\documentclass{aastex}          
\usepackage{spr-astr-addons}    

\begin{document}
%
\title{Globular cluster luminosity function as distance indicator}

\shorttitle{GCLF as Distance Indicator}
\shortauthors{M. Rejkuba}

\author{M. Rejkuba\altaffilmark{1}} 
\email{mrejkuba@eso.org} 

\altaffiltext{1}{ESO, Karl-Schwarzschild-Strasse 2, 85748 Garching b. M\"unchen,
Germany}

Globular clusters are among the first objects used to establish the distance
scale of the Universe. In the 1970-ies it has been recognized that the differential
magnitude distribution of old globular clusters is very similar in different galaxies
presenting a peak  at $M_V \sim -7.5$. This peak magnitude of the so-called 
Globular Cluster Luminosity Function has been then established as a 
secondary distance indicator. The intrinsic
accuracy of the method has been estimated to be of the order of $\sim 0.2$ mag,
competitive with other distance determination methods. Lately the study of 
the Globular Cluster Systems has been used more as a tool
for galaxy formation and evolution, and less so for distance determinations. 
Nevertheless, the collection of homogeneous and large datasets with
the ACS on board HST presented new insights on the usefulness of the 
Globular Cluster Luminosity Function as distance indicator.
I discuss here recent results based on observational and theoretical studies, 
which show that this distance indicator depends on complex physics of the cluster 
formation and dynamical evolution, and thus can have dependencies on
Hubble type, environment and dynamical history of the host galaxy. While the
corrections are often relatively small, they can amount to important systematic
differences that make the Globular Cluster Luminosity Function a less accurate 
distance indicator with respect to some other standard candles.  


%

\section{Introduction}
\label{sect:intro}

Most astrophysics quantities depend critically on accurate distance measurements
and globular clusters are among the first objects used to measure distances.
The literature on their use as distance indicators is very extensive and 
cannot be reviewed fully here. I opted therefore to give a summary on the 
early history of the usage of the globular cluster luminosity function as
distance indicator (Sec.~\ref{sec:history}), before describing the method and 
discussing the way  that globular cluster photometry is used to measure 
distances (Sec.~\ref{sec:method}). 
The main goal of this article is to review the 
recent observational and theoretical results and provide an assessment on the 
accuracy and usefulness of the Globular Cluster Luminosity Function (GCLF) as 
distance determination method. 

Many articles and reviews in the past examined the GCLF as distance indicator
suggesting the accuracy of the method ranging between
0.1--0.2~mag, at the level comparable to, or sometimes even 
better than the other methods used for early type galaxies
 \citep{hanes77a, hanes77b, 
harris+racine79, vandenbergh85, harris91, jacoby+92, ashman+95, kohle+96, whitmore97,
kissler-patig00, harris01}. 
Other works point out more limitations and shortcomings of the 
method \citep[e.g.][]{racine+harris92, tammann+sandage99,ferrarese+00a}. The 
recent review of GCLF as distance indicator by \citet{richtler03} examines in 
particular those galaxies that yield discrepant distance measurements when the GCLF 
distances are compared to those of surface brightness fluctuation method, concluding that discrepant 
cases are most probably caused by peculiar globular cluster populations that are not 
uniformly old, and therefore inapplicable for the method. The most recent 
critical comparison of the relative Virgo vs. Fornax cluster distance as measured using
GCLF and  other standard candles is presented by \citet{villegas+10} and is discussed
briefly here.

\section{A historical note}
\label{sec:history}

%
\begin{figure}[tb]
\includegraphics[width=\columnwidth]{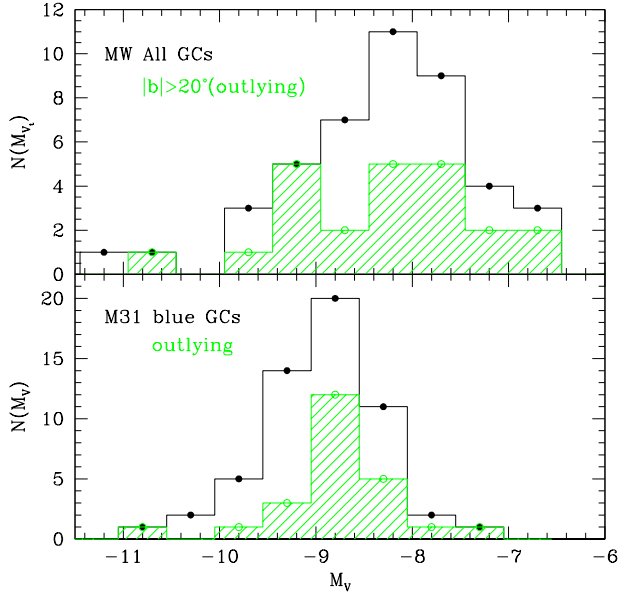}
\caption{The Milky Way (top) and M31 (bottom) globular cluster luminosity functions based on data published by \citet{kron+mayall60} in Tables XV and XX, respectively. In the sample of the Milky Way globular clusters, the absolute magnitudes for the two brightest clusters (47 Tuc and $\omega$ Cen) are taken from \citet{gascoigne+burr56}, as mentioned in a note added in the proofs by \citet{kron+mayall60}} 
\label{fig:KM60gclf}
\end{figure}

Old globular clusters are ubiquitous around all large galaxies. They are compact 
stellar systems containing $10^3$ to $10^6$~M$_\odot$ in stars, and spanning
a range of magnitudes from $M_V \sim 2$ to as bright as $M_V=-11$~mag. Since they
are among the highest luminosity objects in galaxies, they have been recognized
in the early 1960-ies as a potentially important distance indicator, and even before that in 
the second decade of the 20th century Shapley used them to investigate the size of 
the Milky Way galaxy (MW) and our position within it \citep{shapley18a, shapley18b}.

In their 1960 paper on ``Photoelectric photometry of galactic and extragalactic star
clusters" \citet{kron+mayall60} derived the photographic and visual
magnitude distributions of globular clusters in the Galaxy and M31, adopting 
the photometric magnitude $M_{pg}=0.0$ for RR Lyrae variables \citep{sawyerhogg59}. 
The two distributions were found to have a constant difference of 23.6 mag in photographic
light and 24.1 in visual light, which they interpreted as the probable range of
distance moduli for M31. More importantly, the authors concluded that
``the present data do not indicate any substantial difference ... between M31 and
the Galaxy." The luminosity functions of the MW and M31 clusters based on 
data published in that work are shown in Figure~\ref{fig:KM60gclf}. 
The MW GCLF is drawn from Table XV from \citet{kron+mayall60}, augmented
with the measurements from \citet{gascoigne+burr56} for the two most luminous
clusters, 47 Tuc and $\omega$~Cen (the unintentional omission of these two clusters 
is mentioned in a note by \citet{kron+mayall60}). The M31 GCLF is based 
on their Table XX. The absolute magnitudes are derived assuming a distance modulus of 24.6
for M31 \citep{sandage58}. The hatched histograms
show the distribution for outlying clusters, which are presumably less affected by 
rather uncertain extinction corrections. Based on these observations  
\citet{kron+mayall60} derived the mean magnitude of the globular clusters in the
Galaxy of $\langle M \rangle_{\mathrm{Pt}} = -7.7$ and 
$\langle M \rangle_{\mathrm{Vt}} = -8.2$~mag. Applying these values to the
observed mean magnitude for 11 clusters in the Large Magellanic Cloud their
estimated distance modulus for the LMC was $19 \pm 0.5$~mag.

In an independent study of M31 globular clusters \citet{vetesnik62} noticed the
similarity of M31 and MW globular cluster luminosity distributions with a peak at
$M_V \sim -8$~mag. It should be noticed though that these early studies suffer from 
significant biases due to incompleteness and primary distance calibration uncertainties.
In both galaxies the observed clusters are almost all brighter than $M_V \sim -7$. 
In addition, the corrections for often unknown extinction
are not always applied.

Beyond the Local Group one of the first galaxies to have its distance determined
using the observations of its globular clusters is M87. Two studies published 
in 1968 derived values for the Hubble constant based on the derived distance to 
M87, in remarkable agreement with modern measurements made 
within the HST Key Project \citep{freedman+01}. \citet{racine68} 
measured magnitudes of more than 1000 globular cluster candidates around M87 
and derived a distance modulus of $(m-M)_B = 30.7 \pm 0.2$ to M87 based on the 
comparison of the brightest cluster magnitudes with the brightest clusters 
in the MW and M31, as well as based on the peak magnitude of the cluster 
distribution. His  value of the Hubble constant (he actually calls it ``a velocity/distance ratio") 
is $79 \pm 12$ km/sec/Mpc. In a similar study entitled ``A New Determination of the
Hubble Constant from Globular Clusters in M87", published the same year,
\citet{sandage68} derived a distance modulus to M87 of $(m-M)_{AB} = 31.1$, 
based on brightest cluster magnitude, and the Hubble constant of
$75.3^{+19}_{-15}$~km/sec/Mpc.

While these early studies suffered from limitations, the prominent one being 
the inconsistency whether to use the brightest cluster magnitude, the average magnitude of a
sample of globular clusters, or to derive the peak magnitude in the
distribution, they have already in 1960-ies demonstrated that globular clusters may
be useful distance indicators.  Some more modern determinations of the Hubble 
constant using the GCLF are discussed in Section~\ref{sec:hubble}.

In 1970 Racine remarked that the brightest clusters in M87 and in the Local Group 
galaxies, the MW and M31, may  be different \citep{racine70}, and shortly thereafter
the brightest clusters were not used any more as standard candles. Instead the GCLF 
maximum value was established as the standard candle \citep{hanes77a} and 
this method is described in the next section. 

Given that the brightest clusters  are not a standard candle, I do not discuss them 
further here. It is worth mentioning, 
however, that they are well worth a study on their own, because they contain important 
clues for the GCLF formation 
and overall shape \citep[see the discussions by][]{mclaughlin94, richtler03, richtler06}, and 
because they represent a link between the star clusters and the 
so-called Ultra Compact Galaxies or Dwarf-Galaxy Transition Objects 
\citep{hilker+99,hasegan+05,rejkuba+07,mieske+02,mieske+08}.

\section{The GCLF distance determination method}
\label{sec:method}

%
\begin{figure}[tb]
\includegraphics[width=\columnwidth]{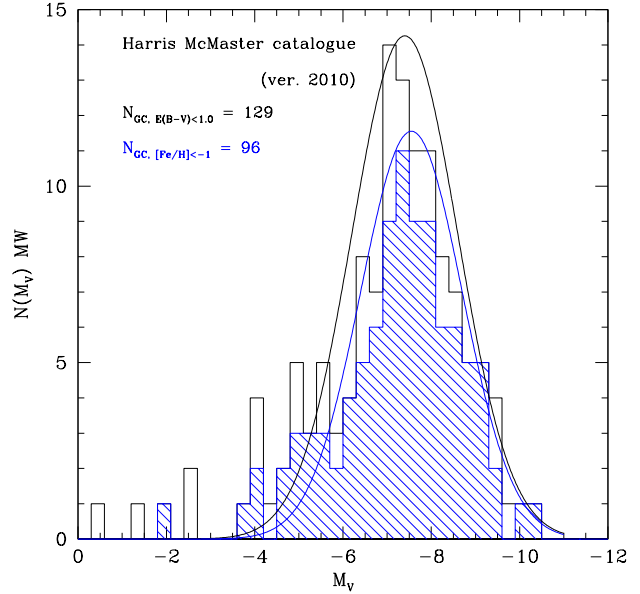} 
\caption{The Milky Way GCLF based on the 2010 version of the \citet{harris96} catalogue for 129
clusters with $\mathrm{E(B-V)} < 1.0$ is shown with black histogram. The subset of these
clusters that have $\mathrm{[Fe/H]}<-1$~dex is plotted as hatched blue histogram} 
\label{fig:MWgclf}
\end{figure}

The main premise of the GCLF method is that the {\it old globular clusters} 
magnitude distribution exhibits a universal shape that is characterised 
by some number that can be used as the standard candle to measure
the distances. The distance determination method is then conceptually very simple
and observationally relatively inexpensive, as it does not require time-series imaging:
it is only necessary to measure luminosities of all globular clusters in a given 
target galaxy reaching a sufficiently deep limiting magnitude to recover the 
characteristic magnitude, build 
its magnitude distribution, and compare it with the known magnitude
distribution of globular clusters in the MW for which the GCLF 
standard  candle is calibrated based on accurate individual cluster 
distances from well understood and well calibrated primary distance indicators. 
Of course, the real measurement has additional complications, and several corrections
and sources of error are discussed below and in Section~\ref{sec:errors}.

Figure~\ref{fig:MWgclf} shows the modern version of the GCLF for 
the MW globular clusters based on the 2010 compilation of the Harris 
McMaster Milky Way Globular Cluster catalogue \citep{harris96}. The black histogram shows the 
distribution of 129 clusters with $\mathrm{E(B-V)} < 1.0$, while a subset of these that
are more metal-poor than $\mathrm{[Fe/H]} < -1$~dex is shown as hatched histogram.
Overplotted to both is a Gaussian fit for the distributions. 
As was already mentioned by other authors \citep[e.g.][]{harris01, richtler03} the name 
Globular Cluster Luminosity Function is a misnomer due to the fact that it refers to the {\it magnitude} and not {\it luminosity} distribution of globular clusters in a galaxy. 

The most commonly adopted way to derive the characteristic magnitude is to 
fit the GCLF with a Gaussian function, as introduced by \citet{hanes77b}:
\begin{equation}
\frac{dN}{dM_V} \propto \frac{1}{\sigma \sqrt{2\pi}} \mathrm{exp}^{- \frac{(M_V-M_{V,0})^2}{2 \sigma^2} }
\end{equation}
The distribution is then characterized by only two parameters: the dispersion (sigma) and the 
peak (or turnover, TO) magnitude. The latter is the standard candle and the distance measurement is done relative to the MW or to the M31 GCLF TO value. The GCLF is a secondary distance indicator, since the absolute distances to the MW and M31 globular clusters must be known. These are based primarily on the RR Lyrae luminosity scale for the MW globular clusters, or alternatively on the Cepheid distance to M31 (Section~\ref{sec:calibration}). 

%
\begin{figure*}[tb]
\includegraphics[width=\columnwidth]{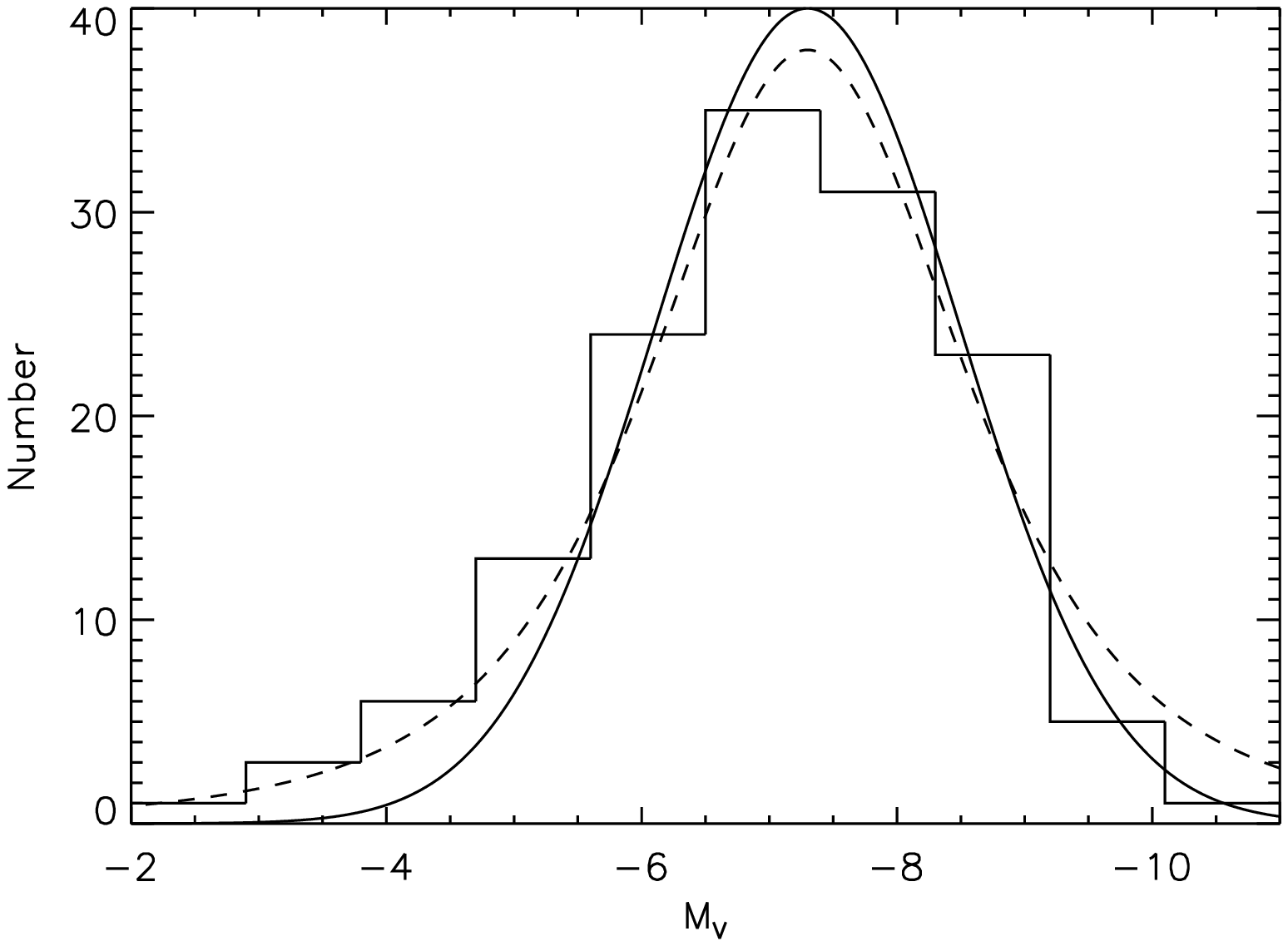} 
\includegraphics[width=\columnwidth]{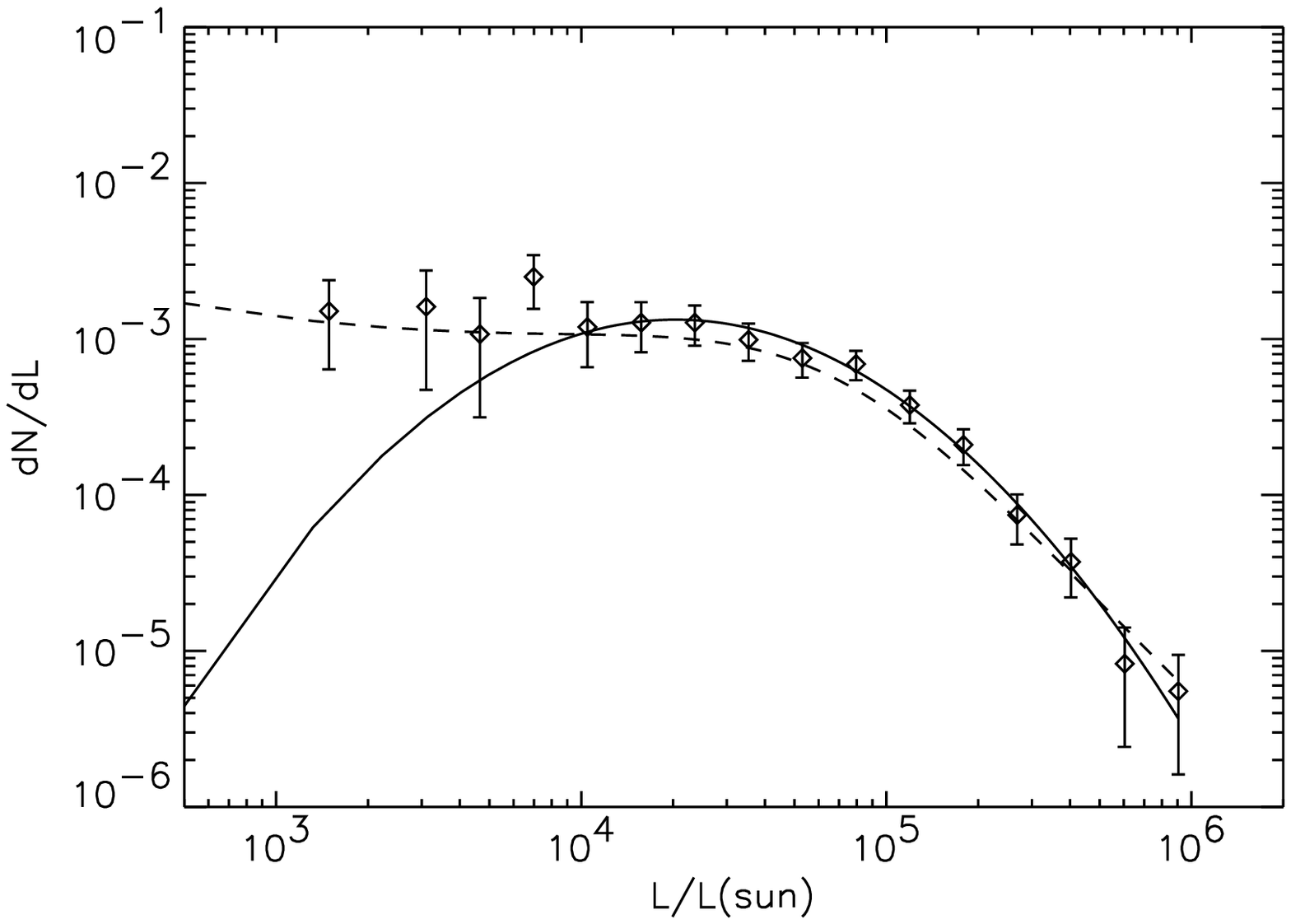} 
\caption{The Milky Way GCLF in magnitude units (left) and in luminosity units (right) fitted with the Gaussian (solid line) and $t_5$ function (dashed line). This figure is from \citet{larsen+01} Figure~1, reproduced by permission of the AAS  } 
\label{fig:larsen01}
\end{figure*}

It has to be emphasised that the fitting of the globular cluster magnitude distribution with a Gaussian provides just a useful parameterisation with an analytic function,  but  it is not  physically motivated \citep{harris01}. This shape of the GCLF is a consequence of the logarithmic magnitude scale combined with a change in power law slope of the underlying distribution of  luminosities of globular clusters, which, combined with the approximately constant mass-to-light ratio for old globular clusters, leads to a power-law mass spectrum of the globular clusters in the form $N(M) \propto M^{-\alpha}$ \citep{harris+pudritz94}.  At the magnitude that corresponds to the peak of the Gaussian the linear luminosity function ($dN/dL \propto L^{-\alpha}$) changes the power-law slope from $\alpha < 1$ to $\alpha > 1$, because then the magnitude distribution exhibits the peak $dN/dm_V \propto 10^{0.4(1-\alpha)m_V} $ \citep{mclaughlin94}. \citet{mclaughlin94} showed that actually the GCLF is non-Gaussian, but as it is observed to be peaked, unimodal and nearly symmetric, its approximation with a Gaussian provides relatively unbiased distance estimate.

Alternative GCLF fitting functional forms for the MW and M31 cluster distributions 
include for example Cauchy and $t$ distribution \citep{secker92}, Gauss-Hermite expansions \citep{abraham+vandenbergh95} and a piecewise exponential fit \citep{mclaughlin94,baum+95}.
\citet{secker92} used a maximum-likelihood estimate to demonstrate that the best fit to the MW and M31 GCLFs is obtained with the $t_5$ distribution in the following form:
\begin{equation}
\frac{dN}{dm} \propto \frac{8}{3 \sqrt{ 5 \pi }\sigma_t } \left(  1 + \frac{(m-m_0)^2}{5 \sigma_t^2} \right)^{-3}
\end{equation}
\citet{kissler+94} also prefered the $t_5$ functional form for the GCLF of NGC 4636, and \citet{larsen+01} used it to fit GCLFs of 17 nearby galaxies. 

The superior fit by the $t_5$ distribution with respect to the Gaussian is due to the asymmetry of the GCLF which presents a longer tail towards the faint end. The more extended faint end of the GCLF is visible in the MW GCLF (Fig.~\ref{fig:MWgclf}) and is even better appreciated from 
Figure~\ref{fig:larsen01}, which is reproduced from  \citet{larsen+01}. The deviation from the symmetry at the faint end of the GCLF is a consequence of the dynamical evolution of the initial cluster luminosity function which is expected to be better represented by a power-law than a 
log-normal distribution \citep[e.g.][]{larsen02}.
In spite of this difference at the faint end, the fit of the GCLF TO yields very similar value irrespective of whether the fit is done using a Gaussian or a $t_5$ function \citep{larsen+01}. 

%
\begin{figure}[tb]
\includegraphics[width=\columnwidth]{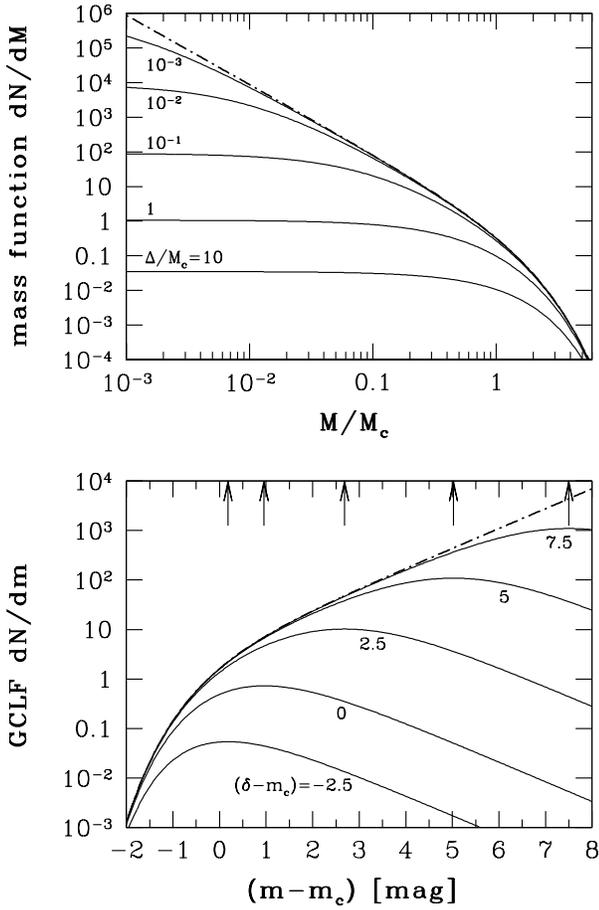} 
\caption{The evolved Schechter mass functions $dN/dM$ are shown in the top panel and the corresponding GCLFs $dN/dm$ are in the bottom panel. This is Figure~1 from \citet{jordan+07}, reproduced by permission of the AAS} 
\label{fig:jordan07}
\end{figure}

Figure~\ref{fig:jordan07} (adopted from \citet{jordan+07}) shows the physically motivated functional form for the evolved Schechter mass functions $dN/dM$ and the corresponding GCLFs $dN/dm$ as proposed by \citet{jordan+07}.  The mass distribution of young star clusters in a galaxy is well described by a power law function $dN / dM \propto M^{-\beta}$ with a characteristic slope of $\beta \sim -2$. At the high mass end a truncation of the star cluster initial mass function is observed \citep[e.g.\ M51:][]{gieles+06a}, \citep[the Antennae:][]{whitmore+99}. This  can be well described with an exponential cut-off similar to what was proposed for the initial luminosity function of globular clusters by  \citet{burkert+smith00}. In a functional form of the \citet{schechter76}  function with a power law index of $\sim 2$ and an exponential cutoff above some large mass scale \citep{gieles+06b} the initial cluster mass function is:
\begin{equation}
\frac{dN}{dM} \propto M_0^{-2} \mathrm{exp}^{-\frac{M_0}{M_c}} .
\label{eq:gcimf}
\end{equation}
Due to dynamical erosion of preferentially low mass clusters the original rising low mass end of the young clusters distribution turns down below the power-law behaviour for clusters with masses smaller than about $2 \times 10^5~M_\odot$, which is the mass of a typical globular cluster at the TO magnitude. Such an evolved Schechter mass function has the form:
\begin{equation}
\frac{dN}{dM} \propto (M + \Delta)^{-2} \mathrm{exp}^{-\frac{M + \delta}{M_c}} 
\end{equation}
where the cumulative mass loss rate is $\Delta = \mu_{ev} t$ such that each cluster has a time dependent mass $M(t) = M_0 - \delta$. 
Relating this function to the GCLF, by assuming a typical mass-to-light ratio for globular clusters, the model GCLF has the following form:
\begin{equation}
\frac{dN}{dm} \propto \frac{10^{-0.4(m-m_c)}}{(10^{-0.4(m-m_c)} + 10^{-0.4(\delta - m_c)} )^2 } \mathrm{exp}^{ \left( -10^{-0.4(m-m_c)} \right)}
\end{equation}
where the magnitude $m=C-2.5 \log m$, $\delta = C-2.5 \log \Delta$, and  
$M_c = C - 2.5 \log M_c$.

\citet{harris+09_coma} note that there is no distinguishable difference between Gaussian and Schechter-like function fits for the measurement of the GCLF TO magnitude. 

\section{Theoretical models}
\label{sec:theory}

Before discussing the calibration of the GCLF TO magnitude, I provide a very brief overview of the disruption processes that erode the initial globular cluster mass function (GCMF) to the present day GCMF that is related to the observed GCLF. 

The dynamical processes that destroy globular clusters in a fixed galactic potential are (i) mass loss due to stellar evolution (this process does not change the shape of the mass function), (ii) dynamical friction ($\tau_{df} \propto M^{-1}$), (iii) tidal shocks heating by passages through bulge or disk ($\tau_{sh} \propto \rho_h P_{cr}$ where $\rho_h$ is the mean density within the half mass radius $R_h$), and (iv) evaporation due to 2-body relaxation ($\tau_{ev} \propto M/\sqrt{ \rho_h}$).  In particular the last two are most important for the transformation of a power law or Schechter-like initial GCMF described by Eq.~\ref{eq:gcimf} \citep{harris+pudritz94,gieles+06a}. 
 
\begin{figure}[tb]
\includegraphics[width=\columnwidth]{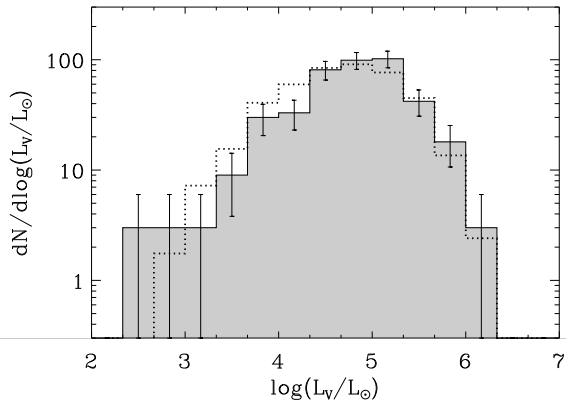} 
\caption{The comparison of the observed (filled) and modeled (dotted) GCLF. This is Figure~3 from \citet{kruijssen+zwart09}, reproduced by permission of the AAS} 
\label{fig:Kruijssen09fig3}
\end{figure}

The pioneering studies of \citet{fall+zhang01,vesperini01,mclaughlin+fall08} successfully reproduced the shape of the GCMF by cluster evaporation. However, the fundamental assumption in those studies is the mass-independent mass-loss-rate, which has been shown to be incorrect \citep{baumgardt+makino03,lamers+05,kruijssen+zwart09}. In particular, \citet{kruijssen+zwart09} demonstrate that due to preferential depletion of low mass stars from globular clusters, the $M/L$ ratio of the clusters changes. They model the influence of the luminosity dependency of $M/L$ on the globular cluster mass function derived from the observed GCLF, providing an improvement to the model of \citet{fall+zhang01}, for the first time fitting the complete GCLF (and the derived GCMF) with a model (Figure~\ref{fig:Kruijssen09fig3}). 
 
%

 An alternative view of the evolution of the globular cluster mass function and evidences for an approximately Gaussian initial mass function of old globular clusters is presented by \citet{parmentier+gilmore07}, who stress the importance of the gas removal process and the protoglobular cloud mass scale.

 The next step is the full blown modeling of the star cluster populations formation and evolution within galaxy simulations, that have changing potential due to their merging history. The first results from \citet{kruijssen+11} provide a better understanding of the cluster evolution and their observed properties.

\section{The calibration}
\label{sec:calibration}

Irrespective whether the GCLF TO is fitted with a Gaussian, $t_5$, or an evolved Schechter function, the fundamental question that needs to be answered before it can be used as a standard candle is whether the GCLF has the same TO peak value in different galaxies. This has been addressed in numerous papers since the early days of the application of the method. Here only few selected recent papers are mentioned in which the TO value was examined and calibrated. 

In their 1992 critical review of selected techniques for extragalactic distance measurements \citet{jacoby+92} concluded that the GCLF method appears to be more accurate than expected, but possibly has a scale error in the sense of overestimating distances by about 13\%. 
That paper emphasised the need to collect high quality observational data for a list of selected large disk and elliptical galaxies close enough to clearly measure their GCLF peak magnitudes in order to provide a robust calibration of the standard candle, as well as the need for a theory of cluster formation and dynamical evolution that can predict the full distribution of globular cluster masses and luminosities.

Roughly at the same time of the review of \citet{jacoby+92}, in the beginning of the 90-ies, a remarkable new discovery in the field of extragalactic globular cluster systems studies was the bimodal color distribution of globular clusters in giant galaxies \citep{zepf+ashman93,ashman+zepf98,brodie+strader06,kissler-patig00}. The presence of the two episodes of vigorous star formation evidenced by the bimodal globular cluster population was quickly recognized as the valuable tool to investigate giant galaxy formation and evolution histories, and as a possible observational evidence of the hierarchical assembly of large galaxies \citep{ashman+zepf92,cote+98}. Therefore from the early 90-ies many studies of the globular cluster systems in external galaxies were concentrating on deriving the color, metallicity, kinematics, specific frequency, sizes and age properties for the two populations, rather than employing them as distance indicators \citep{harris01}. The presence of two globular cluster populations may have brought into question the possibility to obtain a unimodal universal globular cluster luminosity function, and therefore relatively less effort was put in accurate and precise calibrations of the GCLF as distance indicator. 

\citet{whitmore97} reviewed globular clusters as distance indicators concluding that the intrinsic dispersion in $M_{V,TO}$ is $\sim 0.12$~mag for bright ellipticals and providing the mean value of the Gaussian distribution $\langle M_{V,TO} \rangle = -7.21 \pm 0.26$ with the width  $\sigma = 1.35 \pm 0.05$~mag based on literature compilation of GCLF measurements. 

Another extensive discussion of the GCLF calibration by \citet{harris01}, again based on literature selection of reliable GCLF measurements, presented evidence for a small difference between the average TO magnitude for giant ellipticals and disk galaxies. The sample of 16 giant elliptical galaxies has $\langle M_{V,TO} \rangle = -7.33 \pm 0.04$ and mean  $\sigma$ of the Gaussian fit of $\langle \sigma \rangle =  1.4 \pm 0.05$, while in 14 disk galaxies the mean TO parameters are $\langle M_{V,TO} \rangle = -7.46 \pm 0.08$ and  $\langle \sigma \rangle =  1.2 \pm 0.05$. The accuracy of this calibration, however, is limited by the accuracy of the distances to S0 and elliptical galaxies in Virgo, Fornax, and Coma clusters, Leo group and other nearby groups, as well as the accuracy of the primary distance indicators RR Lyrae and HB luminosity for the MW, and Cepheids and possibly the tip of the red giant branch distances to M31.

In the same review \citet{harris01} examines in detail the MW GCLF, and provides, after comparing the MW and M31 GCLFs the TO magnitude of $M_V^0=-7.68 \pm 0.14$ for the MW projected halo sample that is constructed to best represent the globular clusters sample as it would be observed in other galaxies. The TO magnitude for the M31 GCLF halo sample reported in the same review is $M_V^0=-7.80 \pm 0.12$. While the values for these two closest galaxies are consistent within their respective errors, they appear to be significantly brighter than the average for 14 disk galaxies.  Furthermore, the fact that the single review, which admittedly was not focused on the GCLF as the distance indicator primarily, though it did discuss it at a certain length, provided several different values for the TO magnitude of the GCLF, later led to different authors adopting different calibrations in their works. This and other independent calibrations resulted sometimes in different distance measurements based on the same observational datasets.

In addition to the calibration of the TO the question of the universal shape of the whole GCLF was addressed in a number of papers. Most of them compared in some detail the MW and M31 GCLFs and this exercise was repeated as new, deeper data became available \citep[typically data for M31; ][]{vandenbergh85,racine+harris92,ashman+95,barmby+01,huxor+11}. While the result depends on the quality and depth of the photometry, and the sample selection, the Kolmogorov-Smirnov tests show that the M31 GCLF is slightly brighter and narrower than that of the MW, but the differences are typically not statistically significant, and can be explained by a different metallicity of the two populations.

%
\begin{figure}[tb]
\includegraphics[width=4.6cm,angle=270]{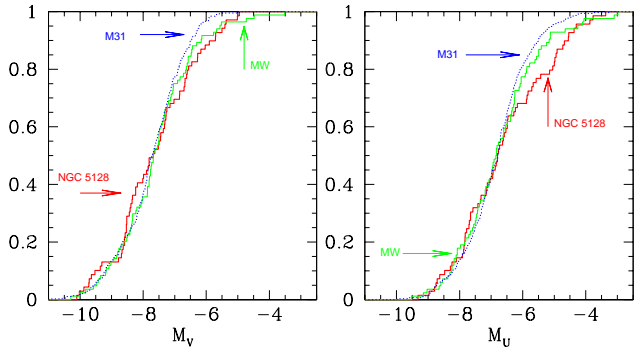} 
\caption{Comparison of the cumulative luminosity function for NGC~5128 (thick red line) with that of the Milky Way (thin green line) and the M31 (dotted blue line) for V (left) and U band (right).
This is Figure~9 from \citet{rejkuba01}, reproduced with permission A\&A \textcopyright ESO} 
\label{fig:Rejkuba01fig9}
\end{figure}

However, the most common application of the GCLF method is to early type galaxies. Elliptical and S0 galaxies are particularly suitable targets because of the large numbers of old globular clusters in their halos. Additionally, in spiral galaxies contamination from the young star cluster populations, as well as crowding and extinction correction problems, are much more severe. Hence for a long time the GCLF distance determination method relied on a leap of faith that the luminosity functions of old globular clusters did not depend strongly on the 
Hubble type. 

The nearest early type galaxy is NGC~5128 \citep{harris+10}, and its GCLF has been sampled to 2.5~mag fainter than the TO (but only in two fields covering a small fraction of the total spatial extent of the galaxy), allowing to demonstrate that the difference between the GCLF of NGC 5128 and the MW is not larger than the difference between the M 31 and the MW GCLFs \citep[see also Fig.~\ref{fig:Rejkuba01fig9}]{rejkuba01}.
In spite of the  significant efforts invested, the very large extension on the sky combined with the need to spectroscopically confirm the cluster candidates due to significant contamination  from the MW foreground stars in typical ground based searches, NGC~5128 still does not have a deep and complete enough GCLF  suitable for an accurate distance determination. Apart from using the globular clusters as galaxy formation and evolution probe, the GCLF in this galaxy may be of particular interest as it could in principle provide an important independent cross-check of many different  distance determinators. NGC~5128 has distances determined from Cepheids \citep{ferrarese+07}, tip of the RGB \citep{harris+99, rejkuba04, harris+10}, planetary nebulae luminosity function \citep{hui+93, harris+10}, Mira period-luminosity relation \citep{rejkuba04} and surface brightness fluctuation method \citep{tonry+01}. In addition novae have been observed in this galaxy \citep{ciardullo+90} and it has even hosted a quite typical type Ia SN \citep{cristiani+92}. Extending the baseline of targets beyond the LMC and M31, in which cross-checks of different distance determination methods can be directly compared, is important.

\citet{dicriscienzo+06} used several recent calibrations of the absolute magnitude of RR Lyrae stars ($M_V(\mathrm{RR})$) to calibrate the TO $V$-band magnitude of the MW GCLF.  The derived peak magnitude depends strongly on the selection criteria for the MW globular cluster sample, as well as, to a lesser extent, on the adopted $M_V(\mathrm{RR}) vs.  \mathrm{[Fe/H]}$ relation. Drawing the full cluster sample from the Harris McMaster Milky Way globular cluster catalogue \citep{harris96} \citet{dicriscienzo+06} derive the peak magnitude $M_V(\mathrm{TO}) = -7.40 \pm 0.09$~mag. This value is identical to the one often quoted as coming from \citet{harris01}. It should be noticed though that the \citet{harris01} review shows the Gaussian curve with $M_V^0=-7.40$ and $\sigma=1.15$ in Figure 1.33, prior to the detailed discussion of the GCLF as distance indicator. In the later sections this review derives $M_V^0=-7.68 \pm 0.14$ for the MW projected halo sample that is constructed to best represent the globular clusters sample as it would be observed in other galaxies. This is very similar to $M_V(\mathrm{TO}) = -7.66 \pm 0.11$~mag derived by \citet{dicriscienzo+06} for a sub-sample of MW globular clusters with $E(B-V) \leq 1.0$ and $2 \leq R_{GC} \leq 35$~kpc. The difference between the TO magnitudes of the two samples is due to differences in their average metallicity (the latter being more metal poor), as well as to environmental effects, since the two cluster groups have a different dynamical history (having been drawn from different parts of the Galaxy). 

%
\begin{figure}[tb]
\includegraphics[width=\columnwidth]{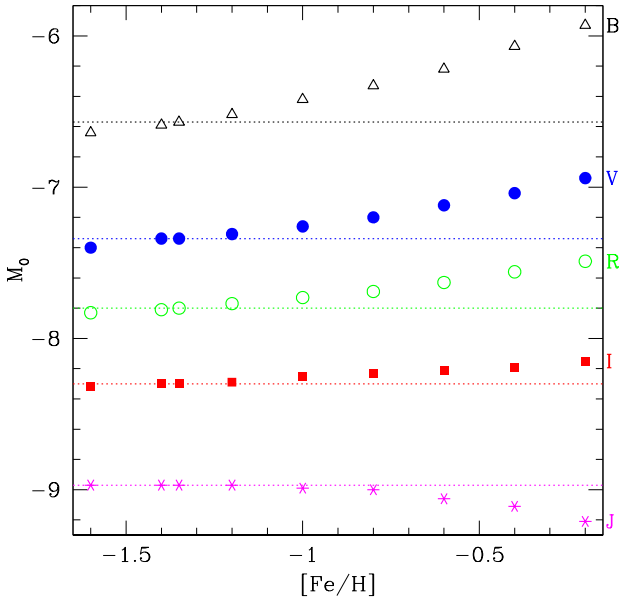} 
\caption{The shift in the peak of the GCLF as a function of metallicity for $B, V, R, I$ and $J$ filters. This figure is based on simulation results reported in Table~3 of \citet{ashman+95} } 
\label{fig:Ashman95_metal}
\end{figure}

The shift in the peak of the GCLF as a function of metallicity for the $B, V, R, I$, and $J$ filters has been investigated by \citet{ashman+95} based on simulations that included the cluster masses drawn from a parent population similar to that of the MW,  and a mean metallicity ranging from $-0.2$~dex to $-1.6$~dex. The GCLF TO in the $V$-band is brighter for more metal-poor globular cluster systems, while in near-IR filters the more metal-poor systems have fainter GCLF TOs. The smallest difference is reported for the $I$-band where a 1.4~dex shift in [Fe/H] results in only 0.17~mag difference in the peak magnitude. Figure~\ref{fig:Ashman95_metal}  based on Table~3 from \citet{ashman+95} summarizes the results of these simulations.  It shows the GCLF TO magnitude shift as a function of metallicity in agreement with the observations. Applying the selection to the metal-poor and metal-rich sample in the MW \citet{dicriscienzo+06}  find  the metal-poor globular clusters sample ($\mathrm{[Fe/H]} < -1.0$, $\langle \mathrm{[Fe/H]} \rangle \sim -1.6$~dex) systematically brighter by $V=0.36$~mag with respect to the metal-rich sample ($\mathrm{[Fe/H]} > -1.0$, $\langle \mathrm{[Fe/H]} \rangle \sim -0.6$~dex), and the same difference is found for M31 clusters \citep{barmby+01}.

\citet{dicriscienzo+06} adopted the same distance modulus to the LMC of 18.50~mag for both the RR Lyrae and the Cepheids distance scales, and used the latter to set the absolute magnitude of the surface brightness fluctuation distance scale. Hence, based on the same LMC fiducial distance scale they compare the TO magnitudes of metal-poor cluster luminosity functions in the MW, M31 and the more distant galaxy sample of \citet{larsen+01}. They find them to be in perfect mutual agreement with $M_V(\mathrm{TO}) = -7.66 \pm 0.11$, $-7.65 \pm 0.19$ and $-7.67 \pm 0.23$~mag, respectively.

\section{Potential for measurement of the Hubble Constant}
\label{sec:hubble}

$H_0$ was derived using the brightest globular clusters as soon the individual cluster magnitudes were measured around M87 in late the 60-ies \citep{racine68, sandage68}.
\citet{vandenbergh+85} used the GCLF TO magnitude of M87 to obtain the Hubble parameter value of $68 \pm 10$~km~s$^{-1}$~Mpc$^{-1}$, although in that work the authors emphasise that the assumption of the invariant GCLF is ``as yet untested". 

Often the main goal of the globular cluster system studies was the investigation of the parent galaxy formation and evolution history, and GCLF distances were done as a secondary use of the data. The uncertainty sources are discussed in the next section, but just to anticipate as stressed by \citet{kissler-patig00}: in particular background/foreground contamination, aperture corrections, incompleteness corrections, GCLF fitting errors, possible contamination from a second younger cluster population, and uncertainty in the absolute calibration of the GCLF TO, were not always fully taken into account, resulting in large scatter and inhomogeneous distances \citep{ferrarese+00b}. Therefore, the compilation of all Hubble constant measurements using the GCLF method is beyond the scope of this review. The GCLF Hubble diagrams based on selected sample of galaxies are shown in reviews by \citet{whitmore97}, \citet{harris01}, and \citet{richtler03}.

Until today, the universality of the GCLF is debated and debatable (see in particular the work of \citet{villegas+10} discussed in the next section).  With deep HST imaging, reaching $V \sim 28$ or fainter, the GCLF method, due to the brightness of its TO magnitude, has the potential to step directly into the Hubble flow. Hence, it can provide measurements of the Hubble constant, without passing through the next distance ladder rung. It can also be used to verify SN~Ia distances independently from Cepheids \citep{dellavalle+98, richtler03}. While most Hubble constant derivations from the GCLF method result in the range of $\sim 62-83$~km/s/Mpc \citep{whitmore+95,sandage+tammann96,whitmore97,harris01,richtler03}, they are primarily based on Virgo cluster (and M87) GCLF distances that still require significant correction for recession velocity. \citet{kavelaars+00} and \citet{harris+09_coma} measured GCLF distances to Coma giant ellipticals obtaining $H_0$ of $69\pm 9$ and $73$~km/s/Mpc, respectively.

\section{Uncertainties}
\label{sec:errors}

Without a discussion of uncertainty sources, this review would be incomplete, though I point out that this topic has been covered very well by reviews of \citet{jacoby+92,whitmore97,harris01} and most recently by \citet{richtler03}. 

%
\begin{figure}[tb]
\includegraphics[width=\columnwidth]{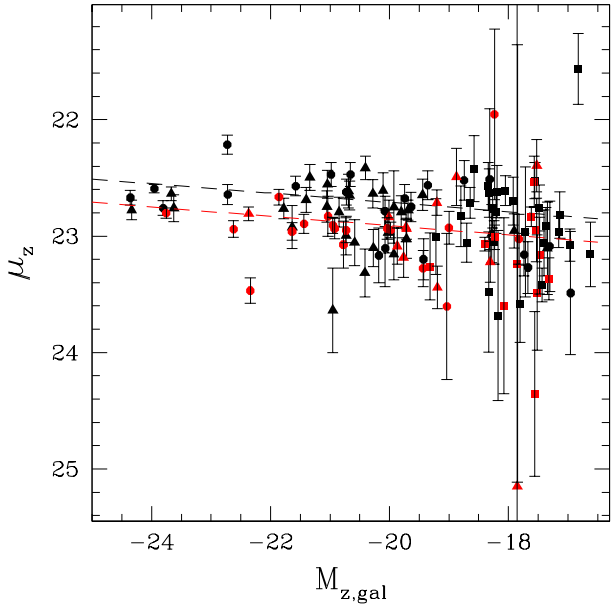} 
\caption{$z$-band GCLF turnover magnitude vs.\ $z$-band galaxy absolute magnitude for all galaxies in the ACS Virgo (black symbols) and ACS Fornax (red symbols) cluster surveys. The lines correspond to simultaneous error-weighted linear fits performed over both samples including an offset of $\Delta(m-M)=0.20 \pm 0.04$~mag, which is the relative distance modulus between the two clusters. This is Figure~6 from \citet{villegas+10}, reproduced by permission of the AAS} 
\label{fig:Villegas10fig6}
\end{figure}

Based on the most recent, thorough and homogeneous study, that used highest quality deep HST observations of 89 early-type galaxies in the Virgo cluster and 43 galaxies in the Fornax cluster, \citet{villegas+10} derive the relative distance modulus between these two galaxy clusters $\Delta(m-M) = 0.20 \pm 0.04$~mag (see also Figure~\ref{fig:Villegas10fig6} adopted from that paper). This value
is systematically 0.22 mag lower with respect to relative distance modulus obtained from  surface brightness fluctuation measurement on the same dataset \citep{blakeslee+09}. Comparison with the relative distance modulus obtained with the planetary nebulae luminosity function (PNLF) method shows also a systematic offset of 0.1-0.15 mag, and the offset with respect to Cepheids is 0.27~mag, again for both in the sense that the relative GCLF distance Virgo-Fornax is smaller than for PNLF and Cepheids \citep{villegas+10}. This points to a systematic error inherent to the GCLF method that makes this distance indicator less accurate with respect to the SBF, PNLF or Cepheid distance indicators.

All possible sources of error are considered here, divided into two groups. 

\subsection{General uncertainty sources}
With high precision, deep CCD photometry, the errors inherent to any photometric method amount to $\sim 0.05$~mag and include the following:
\begin{enumerate}
\item Photometric zero point calibration and photometric magnitude uncertainties for the observed clusters. 

This error depends on the telescope, instrument, and the filter system used. With well understood large telescopes and in particular the stability of the HST, accuracy at the percent level is achievable. Here one should include the aperture correction uncertainty for slightly resolved clusters.

\item Uncertainty associated with extinction correction.

The GCLF method is principally applied to giant elliptical galaxies which have little or no internal extinction, and thus the only important contribution is from the foreground extinction correction error.

\item Systematic errors in the fitting procedure and determination of the GCLF TO magnitude ($m_{0}$).

While the different forms of the fitting functions were discussed above, the fitting method, e.g.\ direct $\chi^2$ fitting or maximum likelihood, can bias the result \citep{secker+harris93}. Other issues to consider here are the photometric limit and completeness corrections, as well as corrections for the foreground contamination by stars and background contamination from compact galaxies masquerading as globular clusters. In galaxies within $\sim 5$~Mpc, under best sub-arcsec seeing conditions, it is possible to distinguish globular clusters from foreground stars due to their slightly resolved profiles \citep{rejkuba01}, while HST can resolve clusters up to about 40~Mpc \citep{kundu+whitmore01a,kundu+whitmore01b,jordan+09, harris09}. More distant galaxies need statistical subtraction of the contaminants. The size and quality of this correction depends on the filter system used \citep{dirsch+03}, on the (Galactic) coordinates of the target, and the availability of the suitable, equally deep, comparison background field. 

\end{enumerate}

\subsection{GCLF specific uncertainty sources}
The errors particular to the GCLF method amount to $\sim 0.3$~mag and include the following:
\begin{enumerate}
\item Uncertainty in the primary calibrator luminosity -- distance scale.

The individual globular clusters in the MW, the primary GCLF calibrator, are 
based on primary distance indicators such as RR Lyrae, horizontal branch, main 
sequence, sub-dwarf, or white dwarf sequence fitting, and astrometric distances 
\citep[e.g.][]{rees96,gratton+99,zoccali+01,richer+04,recio-blanco+05,coppola+11}. 
For M31, an independent GCLF calibration is possible using Cepheids and the tip 
of the red giant branch.
Their accuracy is thought to be at the level of $\sim 10$\% (for details I refer to 
other reviews in this volume).

\item Intrinsic dispersion of $m_0$ and the dependence on the globular cluster sample.

\citet{villegas+10} demonstrate that the intrinsic dispersion of $m_0$ is 0.21 mag for the Virgo cluster and 0.15~mag for Fornax.

\item Environment and dynamical evolution.

This is the dominant uncertainty of the method. \citet{villegas+10} find a systematically lower relative 
distance between Fornax and Virgo clusters using GCLF when compared to SBF, PNLF, and Cepheid distances. This may imply that galaxy clusters with higher velocity dispersion (higher masses) host galaxies with fainter TOs in their globular cluster systems as suggested by \citet{blakeslee+tonry96}.

\item $m_0$ dependence on the Hubble type or galaxy luminosity.

\citet{harris01} finds a dependence on the Hubble type at the level of $\sim 0.13$~mag, while \citet{villegas+10} measure systematically fainter GCLF TO for low luminosity galaxies. Dwarf galaxies in Virgo and Fornax clusters show systematically $\sim 0.3$~mag fainter GCLF TOs, and less populated globular cluster systems. Hence the GCLF method is not applicable to dwarf galaxies fainter than $M_B \sim -18$.  It is thus very surprising that the GCLF TO  of the {\it combined sample} of 175 globular cluster candidates in 68 faint ($M_V>-16$~mag) dwarf galaxies is $M_{V,TO}=-7.6 \pm 0.11$~mag \citep{georgiev+09}, very similar to the metal-poor GCLF TO of the MW.

\item Corrections due to a different metallicity of the target galaxy and the calibrator globular cluster samples.

The bimodal globular cluster color distribution is present in most giant galaxies.  
The metal-poor, blue, population has been shown to be old in essentially all well studied galaxies, while the red population can be either old and metal-rich or contain a somewhat younger cluster population \citep{brodie+strader06}. All galaxies contain the blue peak in the globular cluster color distribution with a relatively shallow dependence of the mean color of that peak on the galaxy luminosity, while the fraction of red clusters and their mean color decreases much faster for fainter galaxies \citep{peng+06}.  The ubiquitous old metal-poor population can be used to minimize the corrections due to metallicity differences. This has been advocated also in the review by \citet{kissler-patig00}.

For example, just selecting metal-poor cluster populations in the MW, M31 and other galaxies \citet{dicriscienzo+06} find $M_{V,TO}$ values consistent to within 0.02~mag for the three samples. In addition as already mentioned in the previous section the use of near-IR filters minimize metallicity corrections \citep{ashman+95}.

It should be noted here that \citet{villegas+10} use full globular cluster samples in each galaxy. Restricting the GCLFs to only the blue population might remove a possible bias due to different average metallicity of galaxies in the two clusters, but this correction is expected to be anyway small due to the near-IR $z$ filter used in this study.

\item The uncertainty in the age of the cluster sample.

The method only works for the {\it old} globular clusters that had undergone the Hubble time of stellar and dynamical evolution in the potential of large elliptical galaxies. The younger star cluster systems will display a fainter TO. This is somewhat counter-intuitive, as the younger clusters are brighter. However, the fainter TO is due to GCLF peak having lower mass clusters and thus fainter average magnitude.

The discrepant, 0.22~mag smaller, relative distance modulus between Fornax and Virgo clusters from the GCLF method could be explained by $\sim 3$~Gyr younger globular cluster systems in Fornax cluster with respect to Virgo \citep{villegas+10} galaxies. Such a trend in globular cluster formation time is not unexpected based on hierarchical galaxy formation simulations \citep{delucia+06}, but again this points to a systematic dependence on environment, which makes the GCLF method a less accurate distance indicator.

\end{enumerate}
 
\section{Summary}
\label{sec:conclude}

The globular cluster luminosity function (GCLF) is an attractive distance indicator for early-type giant galaxies. Through a simple functional form the characteristic turnoff magnitude of the GCLF can be easily obtained and its intrinsic luminosity enables distance measurements up to $\sim 120$~Mpc, to the Coma cluster, and therefore direct determination of the Hubble constant.

It is recommended to apply the method to the metal-poor globular cluster sub-population and to use near-IR filters, as they provide least corrections for metallicity differences between cluster samples. The calibration of the $V$ band TO magnitude for metal-poor clusters in the MW, M31 and 17 other nearby galaxies shows remarkable consistency within only $0.02$~mag, and the weighted mean value is $M_{V,TO} = -7.66 \pm 0.09$, calibrated with respect to the LMC distance modulus of 18.50 \citep{dicriscienzo+06}. 

The accuracy of the method is impacted however by a potential dependency of the standard candle
on environment as it provides a systematically shorter relative distance, by about 0.1-0.27~mag, between the Virgo and Fornax clusters when compared with their relative distances from the PNLF, SBF and Cepheid distance indicators \citep{villegas+10}.

%
\acknowledgments
I thank Soeren Larsen, Andr\'es Jord\'an, Diederik Kruijssen, and Daniela Villegas for permission to reproduce figures from their papers.  Permissions of the AAS and A\&A to reproduce material from their journals  is acknowledged. I am grateful to Michael Hilker, Bill Harris, Tom Richtler, and Markus Kissler-Patig for careful reading and many comments which improved this paper. 
This paper originates from the invited review at the conference "Fundamental Cosmic Distance Scale: State of the Art and the Gaia Perspective". I thank the conference organizers for a very enjoyable meeting and for financial support. 


%
\bibliographystyle{spr-mp-nameyear-cnd}  
\bibliography{/Users/mrejkuba/Work/publications/Article/mybiblio.bib}                

%

\end{document}